\documentclass[preprint,preprintnumbers,amsmath,amssymb,superscriptaddress]{revtex4}
\usepackage{txfonts}
\usepackage{graphicx}
\usepackage{dcolumn}
\usepackage{bm}
\usepackage{array}
\usepackage[colorlinks=true,plainpages=false,linkcolor=blue,urlcolor=blue,citecolor=blue,pdfpagemode=UseNone,pdfstartview=FitBH]{hyperref}

\begin{document}

\title{ Microscopic evidence for the intra-unit-cell electronic nematicity inside the pseudogap phase in YBa$_2$Cu$_4$O$_8$}
\author{W. Wang}
\affiliation{Institute of Physics, Chinese Academy of Sciences,\\
 and Beijing National Laboratory for Condensed Matter Physics,Beijing 100190, China}
\affiliation{School of Physical Sciences, University of Chinese Academy of Sciences, Beijing 100190, China}

\author{J. Luo}
\affiliation{Institute of Physics, Chinese Academy of Sciences,\\
 and Beijing National Laboratory for Condensed Matter Physics,Beijing 100190, China}

\author{C. G. Wang}
\affiliation{Institute of Physics, Chinese Academy of Sciences,\\
 and Beijing National Laboratory for Condensed Matter Physics,Beijing 100190, China}
\affiliation{School of Physical Sciences, University of Chinese Academy of Sciences, Beijing 100190, China}

\author{J. Yang}
\affiliation{Institute of Physics, Chinese Academy of Sciences,\\
 and Beijing National Laboratory for Condensed Matter Physics,Beijing 100190, China}

\author{Y. Kodama}
\affiliation{National Institute of Advanced Industrial Science and Technology (AIST), Umezono, Tsukuba 305-8568, Japan}

\author{R. Zhou}
\thanks{Electronic address: rzhou@iphy.ac.cn}
\affiliation{Institute of Physics, Chinese Academy of Sciences,\\
 and Beijing National Laboratory for Condensed Matter Physics,Beijing 100190, China}
\affiliation{Songshan Lake Materials Laboratory, Dongguan, Guangdong 523808, China}

\author{Guo-qing Zheng}
\affiliation{Institute of Physics, Chinese Academy of Sciences,\\
 and Beijing National Laboratory for Condensed Matter Physics,Beijing 100190, China}
\affiliation{School of Physical Sciences, University of Chinese Academy of Sciences, Beijing 100190, China}
\affiliation{Department of Physics, Okayama University, Okayama 700-8530, Japan}

\date{\today}

\begin{abstract}
{Understanding the nature of the mysterious pseudogap phenomenon is one of the most important issues associated with cuprate high-$T_c$ superconductors.
Here, we report $^{17}$O nuclear magnetic resonance (NMR) studies on two planar oxygen sites in stoichiometric cuprate YBa$_2$Cu$_4$O$_8$ to investigate the symmetry breaking inside the pseudogap phase. We observe that the Knight shifts of the  two oxygen sites are identical at high temperatures  but different below $T_{\rm nem} \sim$ 185 K, which is close to the pseudogap temperature $T^{\ast}$. Our result provides a microscopic evidence  for intra-unit-cell electronic nematicity. The difference in quadrupole resonance frequency between the two oxygen sites is unchanged below $T_{\rm nem}$, which suggests that the observed nematicity does not directly stem from the local charge density modulation.
Furthermore,  a short-range charge density wave (CDW) order is observed below $T \simeq$ 150 K. The additional broadening in the $^{17}$O-NMR spectra because of this CDW order is determined to be inequivalent for the two oxygen sites, which is similar to that observed in case of nematicity.
These results suggest a possible connection between nematicity, CDW order, and  pseudogap.}
\end{abstract}



\maketitle


\section{Introduction}\label{section1}

The origin of the pseudogap  is essential for understanding the mechanism of high-$T_c$ superconductivity in cuprates.
Various orders have been found within the pseudogap phase.
One major issue is to identify the reported orders (broken symmetries) inside the pseudogap phase and elucidate the manner in which they are interrelated\cite{Keimer2010,Fradkin2010}.
Recently, a nematic state, i.e., an electron-correlated state exhibiting spontaneous rotational symmetry breaking, has become an emerging research topic\cite{Kohsaka2007,Lawler2010,Mesaros2011,Wu2017,Auvray2019}.
In particular, in the most studied cuprate system, YBa$_2$Cu$_3$O$_y$ (YBCO), the breaking of the $C_4$ symmetry was identified via various bulk measurements. The spin structure factor centered at the N\'{e}el ordering vector becomes elliptical\cite{Hinkov2008}, and the in-plane Nernst effect and magnetic susceptibility become anisotropic\cite{Daou2010,Sato2017} below the onset temperature $T_{\rm nem}$, which  coincides with the pseudogap opening temperature $T^\ast$.

However,
the microscopic origin of nematicity 
has not yet been understood. Further, it has not yet been determined whether the observed nematicity is an intrinsic phenomenon of the CuO$_2$ plane because the crystal structure of YBa$_2$Cu$_3$O$_y$ breaks the $C_4$ symmetry because of the presence of the CuO chain. Even if nematicity is intrinsic to the CuO$_2$ planes, it has not yet been determined whether it is a consequence of the symmetry breaking field with large nematic susceptibility or a long-range  ordered state.
In addition, although the scaling behavior of the nematic order parameter was observed
\cite{Sato2017,Shekhter2013},
the question whether the $C_4$ symmetry breaking is the primary cause or
secondary effect of the pseudogap still remains.
This is considerably important because nematic quantum fluctuation was recently proposed as a new candidate for the pairing mechanism and strange metallic behavior in cuprates\cite{Badoux2016,Lederer2017}.
Furthermore, the relation between nematicity and other orders, such as antiferromagnetic order\cite{Kee09} and charge density wave order\cite{Wu2011,Ghiringhelli2012,Gerber2015,Chang2016}, remains unclear.
To resolve these issues, the microscopic studies on a suitable system are needed.

NMR is a powerful local probe for studying nematicity, as demonstrated in Fe-based superconductors\cite{Fu2012,Baek2015,Iye2015,Zhou2016,Dioguardi2016}. The oxygen-stoichiometric YBa$_2$Cu$_4$O$_8$ is a unique cuprate superconductor with the same bilayer CuO$_2$ planes as YBa$_2$Cu$_3$O$_y$ but with double fully-filled CuO chains.
In contrast to YBa$_2$Cu$_3$O$_y$\cite{Wu2016},
as will be shown later, no additional NMR lines related to fractional oxygen stoichiometry in the chains are observed, indicating that such kind of disorders is almost absent.
This makes it an ideal compound to study the nature of nematicity and CDW and their relation with the pseudogap.

In this study, we report the result of $^{17}$O-NMR measurements on YBa$_2$Cu$_4$O$_8$ to shed new light on the nature of nematicity. The Knight shift at the two planar oxygen sites becomes inequivalent below $T_{\rm nem} \sim$ 185 K, which is close to the pseudogap temperature $T^\ast$. The difference in quadrupole frequency between the two sites remains unchanged below $T_{\rm nem}$. These results provide microscopic evidence of intra-unit-cell electronic nematicity and indicate that it  is not directly related to the local charge density between the two oxygen sites.
We further revealed a short-range nematic CDW order below $T_{\rm CDW} \sim$ 150 K, below which the $^{17}$O-NMR linewidth is anomalously broadened.
Our observations indicate that nematicity and short-range CDW order are intrinsic to the CuO$_2$ plane and are likely linked to each other.

\begin{figure}
\includegraphics[width= 8 cm]{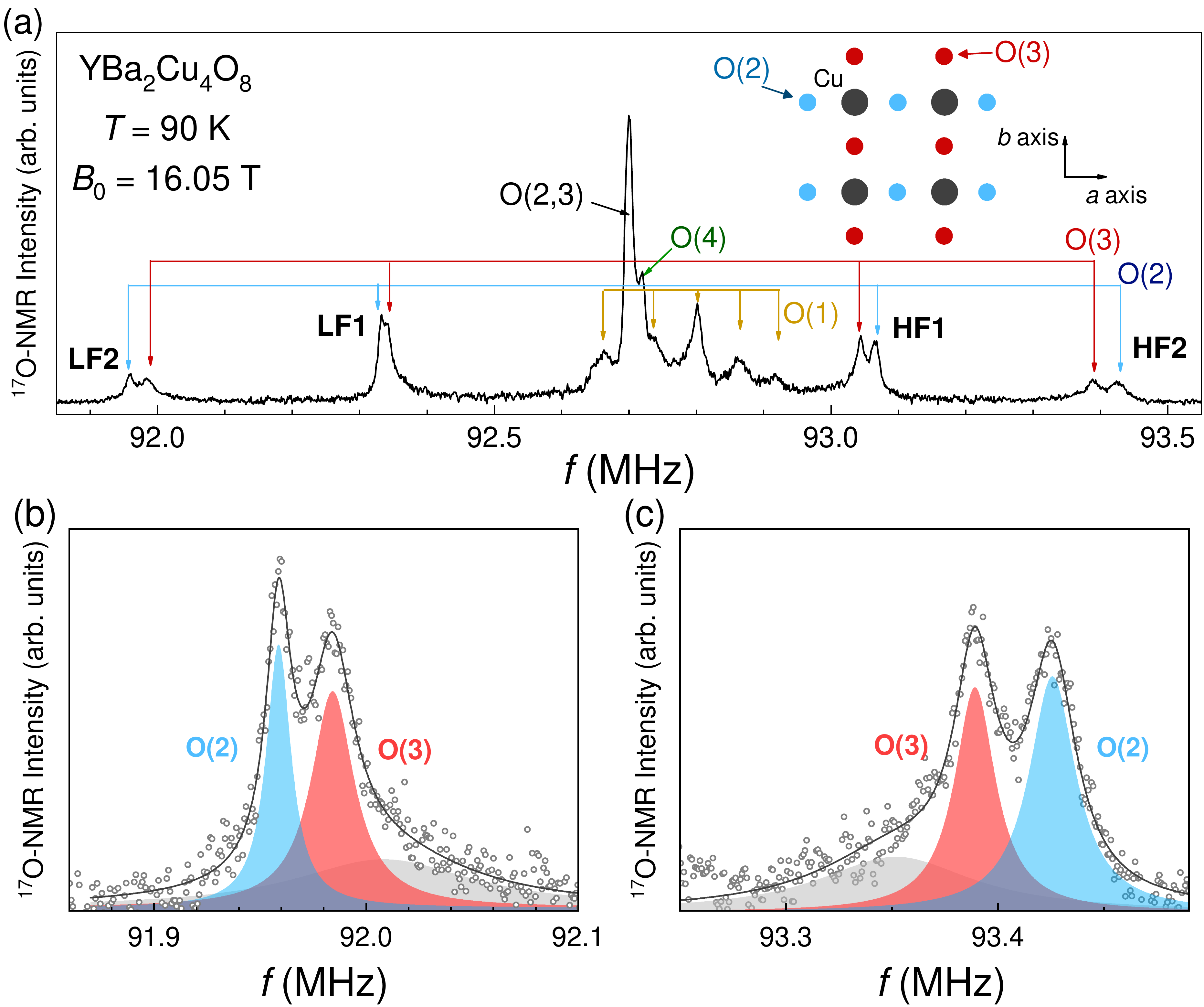}
\caption{(Color online) (a) Typical $^{17}$O-NMR spectrum of YBa$_2$Cu$_4$O$_8$ at $T$ = 90 K and $B_0$ = 16.05 T applied parallel to the $c$-axis. O(2) and O(3) are the sites in CuO$_2$ planes lying in bonds oriented along the crystalline $a$ and $b$ axes, as shown in the insert.
HF1 (HF2) represents the first (second) high-frequency satellites, and LF1 (LF2) represents the first (second) low-frequency satellites.
The five small peaks close to the O(2,3) central peak correspond to the O(1) site of CuO chains, as labeled using yellow arrows. The central line from the apical oxygen site is also observed, as labeled by O(4). (b) and (c) are the magnified view of LF2 and HF2 with fits by three Lorentzian functions. The small gray peak is from the misaligned part of the powder sample.
}
\label{spec}
\end{figure}


\section{Experimental results}\label{section2}

The polycrystalline sample of YBa$_2$Cu$_4$O$_8$ used for NMR measurements was grown under high oxygen pressure\cite{Kodama1990}. The powder sample was magnetically aligned along the $c$-axis. Details regarding sample synthesis, $^{17}$O substitution, and magnetic alignment can be found in ref. 26. The superconducting transition temperature $T_c$ of our sample is 80 K, as measured by AC  susceptibility\cite{SM}, which is among the highest values for this compound, indicating its high quality.
We identify doping in YBa$_2$Cu$_4$O$_8$ in the same way as in YBa$_2$Cu$_3$O$_y$\cite{Liang2006}, namely $p$ = 0.135.
The $^{17}$O-NMR spectra were obtained by adding Fourier transforms of the spin-echo signal recorded for regularly spaced frequency values.

\subsection{Intra-unit-cell nematicity}

Figure \ref{spec} shows the typical NMR spectrum in the field parallel to the $c$-axis, where five sets of peaks originate from different nuclear transitions of the nuclear spin $I$ = $5/2$.
Two peaks are clearly observed for all the satellites; these peaks correspond to the planar oxygen sites, O(2) and O(3), lying in bonds oriented along the crystalline $a$ and $b$ axes, respectively.
Because of the presence of CuO chains, the electric field gradients (EFG) at the O(2) and O(3) nuclear sites are not identical. Thus, the separation between the adjacent quadrupole satellites, i.e., nuclear quadrupole frequency $\nu_c$, is different between the two planar oxygens. In YBa$_2$Cu$_3$O$_y$, the $\nu_c$ of O(2) is determined to be larger than that of O(3) by performing the experiment on de-twined single crystals\cite{Vinograd2019}. In the same way, we assign O(2) and O(3), as labeled in Fig. \ref{spec}.

\begin{figure}
\includegraphics[width= 8 cm]{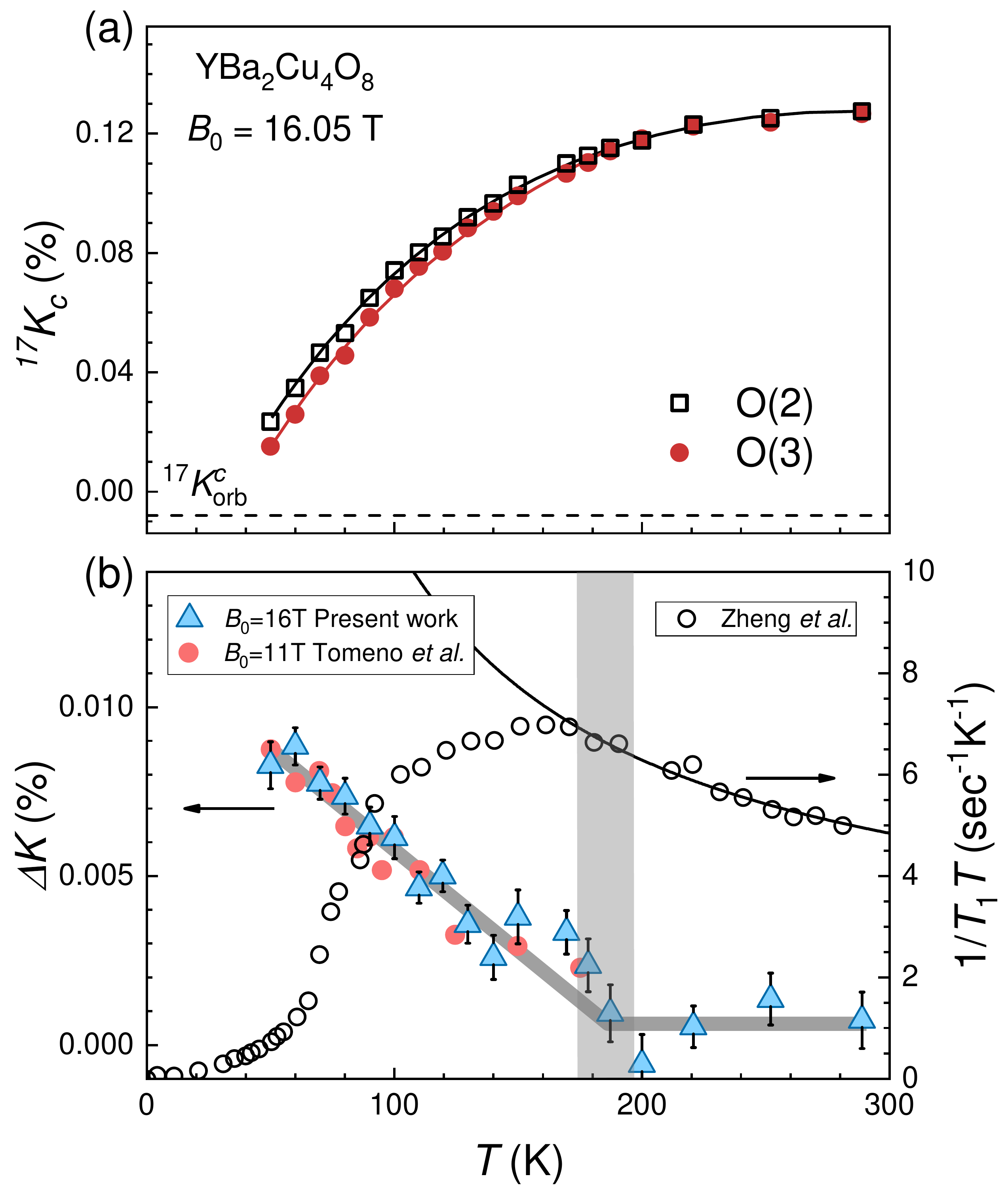}
\caption{(Color online) (a) $T$-dependent Knight shift of the O(2) and O(3) sites. The dashed line indicates the value of $K_{orb}$ = -0.008 $\%$\cite{Mangelschots1992}. Solid lines are the guide to the eye. The error bars are the standard deviation in fitting the NMR spectra, as shown in Fig. \ref{spec} b and c, and are smaller than the symbol size. (b) Superposition of ${\Delta}K = K_{c}^{\text{O(2)}} - K_{c}^{\text{O(3)}}$ and 1/$T_1$$T$ of the planar Cu site as a function of temperature. The red circles are from the previous report measured at 11 T\cite{Tomeno1994}. The 1/$T_1$$T$ result is from a previous study that used NQR\cite{Zheng1999}. The solid line is a fit of 1/$T_1T$ above $T$ = 150 K obtained using the Curie-Weiss law as (1/$T_{1}T$)=$a$+$b$/($T$+$\theta$).}
\label{shift}
\end{figure}

\begin{figure}
\includegraphics[width= 8 cm]{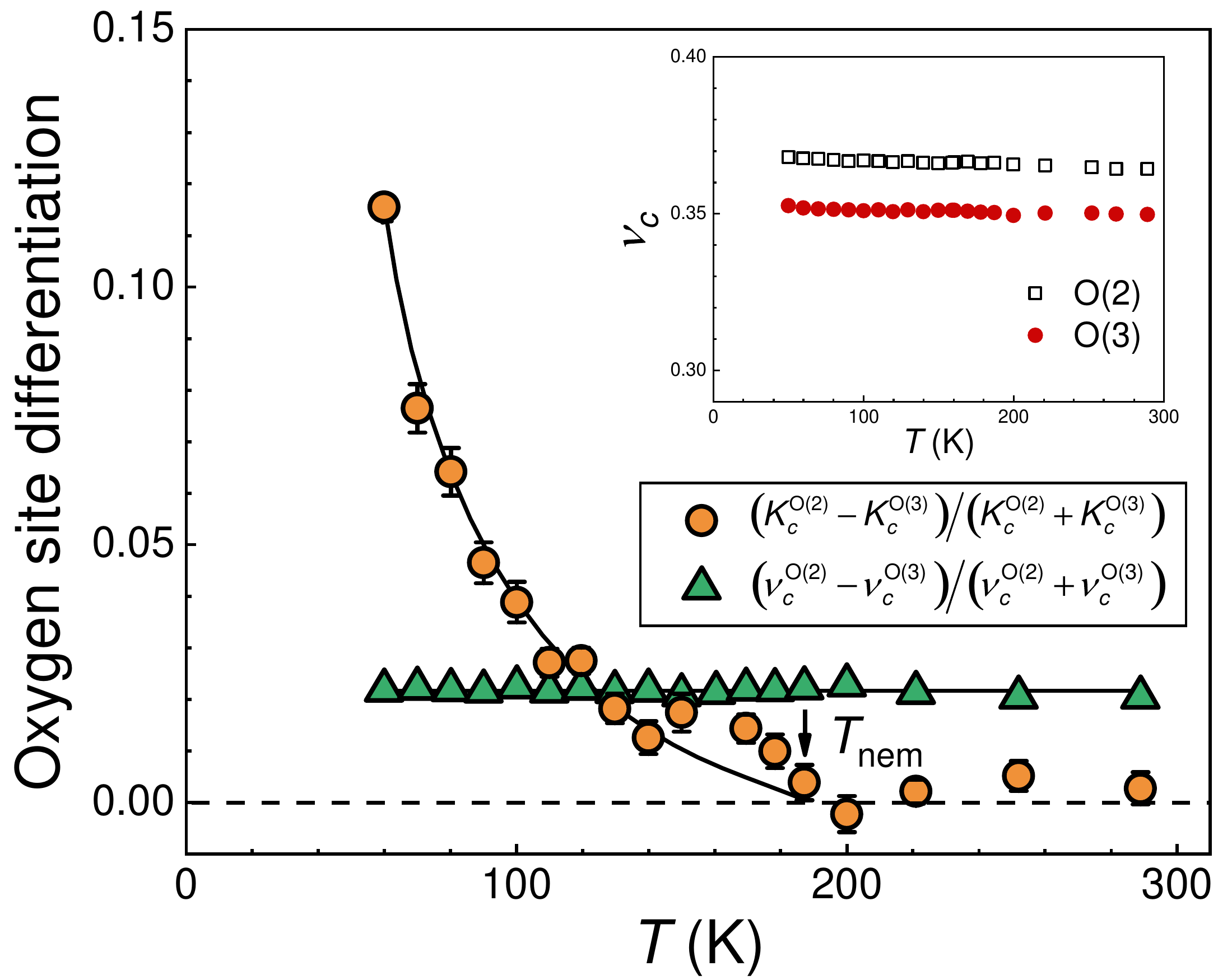}
\caption{(Color online) Comparison of the oxygen site differentiation in the Knight shift defined as $\left( K_{c}^{O(2)}-K_{c}^{O(3)} \right)/\left( K_{c}^{O(2)}+K_{c}^{O(3)} \right)\ $ with that in the quadrupole frequency $\nu _{c}$ defined as ${\left( \nu _{c}^{O(2)}-\nu _{c}^{O(3)} \right)}/{\left( \nu _{c}^{O(2)}+\nu _{c}^{O(3)} \right)}\;$. The inset shows the $T$-dependence of $\nu _{c}^{\text{O(2)}}$ and $\nu _{c}^{\text{O(3)}}$. Solid lines are the guide to the eye. }
\label{asy}
\end{figure}

The separation between O(2) and O(3) in case of the first low-frequency satellite (LF1) is much smaller than the linewidth below 150 K; thus, the two sites cannot be easily distinguished. Therefore, we focus on the second high-frequency (HF2) and second low-frequency (LF2) satellite lines, as shown in Fig. \ref{spec} (b) and (c). For all the satellite lines, a tail is observed toward the central line position. This occurs because a small portion of the sample is not aligned well. We fitted the peaks using three Lorentzian functions, as shown by the solid areas in Fig. \ref{spec} (b) and (c).
$K_c$ is defined as ${{K}_{c}}=\left( f-{{f}_{2\text{nd}}}-\gamma {{B}_{0}} \right)/\gamma {{B}_{0}}$. Here, $f$ is the averaging resonance
frequency of LF2 and HF2 and $\gamma$ is the gyromagnetic ratio of $^{17}$O. ${f}_{2\text{nd}}$ is the second-order quadrupole shift, which is very small in case of high fields and $T$-independent because the $\nu_c$ values of O(2) and O(3) are nearly $T$-independent, as shown in the inset of Fig. \ref{asy}.
Therefore, ${f}_{2\text{nd}}$ will not be discussed further. In Fig. \ref{shift} (a), we plot the temperature dependence of $K_c$ for O(2) and O(3).
At high temperatures, $K_{c}^{\text{O(2)}}$ and $K_{c}^{\text{O(3)}}$ are identical. They begin becoming inequivalent below $T \sim$ 185 K. Such tendency can also be inferred from a previous study at a lower field of $B_0$ = 11 T\cite{Tomeno1994}. In Fig. \ref{shift} (b), we show a comparison of the temperature dependence of ${\Delta}K = K_{c}^{\text{O(2)}} - K_{c}^{\text{O(3)}}$ between our work and a previous study\cite{Tomeno1994}. The two sets of data with different applied fields are entirely consistent with each other. This indicates that ${\Delta}K$ is field independent and, thus, cannot be ascribed to the second-order quadrupole shift.
Furthermore, we compare ${\Delta}K$ and 1/$T_1$$T$ of planar Cu in Fig. \ref{shift} (b)\cite{Zheng1999}.
Previously, in the NMR community, the pseudogap opening temperature was defined as $T^{\ast}_{\rm max} \sim$ 150 K at which 1/$T_1$$T$ is the maximum\cite{Zheng1999,Bankay1994}. The $T^{\ast}_{\rm max}$ obtained in this manner is lower than the pseudogap temperature $T^{\ast} \sim$ 170 K obtained through transport measurements\cite{Hussey1997}. 1/$T_1$$T$ at high temperatures is related to the antiferromagnetic spin fluctuations; thus, the 1/$T_1$$T$ above 150 K is fitted using a Curie-Weiss function as $a$ + $b$/($T$+$\theta$). Then one can also choose the onset temperature $T^{\ast}_{\rm CW}$ of the pseudogap opening at which 1/$T_1$$T$ starts to deviate from the Curie-Weiss function, which is around 180 K as shown in Fig. \ref{shift} (b).
In any case, we find that the onset temperature of ${\Delta}K$, which becomes nonzero, is very close to the pseudogap temperature $T^{\ast}$, indicating that the Knight shift of the two planar oxygen sites is inequivalent in the pseudogap phase.

One apparent possibility for the non-zero ${\Delta}K$ is related to the orthorhombic structure or the CuO chains, which can affect the local properties of the planar oxygen sites,
but such effect should not have a temperature dependence.
However, the discrepancy between $K_{c}^{\text{O(2)}}$ and $K_{c}^{\text{O(3)}}$ defined as $\left( K_{c}^{O(2)}-K_{c}^{O(3)} \right)/\left( K_{c}^{O(2)}+K_{c}^{O(3)} \right)\ $ exhibits a strong temperature dependence below $T^\ast$, as shown in Fig. \ref{asy}. These results indicate that the $T$-dependence of ${\Delta}K$ is $not$ related to the orthorhombic structure or CuO chains. The $\nu_c$ values of O(2) and O(3) are also different; however, the discrepancy between $\nu _{c}^{\text{O(2)}}$ and $\nu _{c}^{\text{O(3)}}$ defined as ${\left( \nu _{c}^{O(2)}-\nu _{c}^{O(3)} \right)}/{\left( \nu _{c}^{O(2)}+\nu _{c}^{O(3)} \right)}\;$ is $T$-independent, as shown in Fig. \ref{asy}, which is an effect of the orthorhombic structure or the presence of CuO chains. 

We note that the difference in hyperfine coupling between O(2) and O(3) because of different local crystal environments would not be able to account for the observed ${\Delta}K$, as elaborated below. The Knight shift $K_c$ is related to the static, uniform, magnetic susceptibility as ${{K}_{c}}=K_{c}^{\text{spin}}+K_{c}^{\text{orb}}=A_{c}^{\text{spin}}\chi _{cc}^{\text{spin}}+A_{c}^{\text{orb}}\chi _{cc}^{\text{orb}}$.
Here, $\chi_{cc}^{\text{spin}}$ and $\chi_{cc}^{\text{orb}}$ are the spin and orbital (Van Vleck) susceptibility, respectively. $A_{c}^{\text{spin}}$ and $A_{c}^{\text{orb}}$ are the hyperfine constants related to the spin and orbital susceptibility, respectively.
If one assumes that the ${\Delta}K$ at low temperatures is due to ${\Delta}K$ =($A^{\rm O(2)}$ - $A^{\rm O(3)}$)$\chi$ = ${\Delta}A$ $\chi$, then ${\Delta}K$ at $T$ = 280 K will be nearly five times larger than that at $T$ = 50 K because $\chi$ is larger at high temperatures. However, ${\Delta}K$ = 0 for $T$ $>$ 180 K was observed experimentally. One might further argue that ${\Delta}A$ can be temperature dependent. However, to be fully consistent with the observed ${\Delta}K$, the anisotropy between $A^{\rm O(2)}$ and $A^{\rm O(3)}$, ($A^{\rm O(2)}$ - $A^{\rm O(3)}$) / ($A^{\rm O(2)}$ + $A^{\rm O(3)}$), would need to change from 0 at 280 K to 0.15 at 50 K. Such a large change is impossible because the anisotropy of the in-plane lattice parameters, $(a - b)/(a + b)$, only changes by a few percent between 280 K and 50 K\cite{Alexandrov1990}.
Finally, the difference in $K_{\rm orb}$ between O(2) and O(3) cannot explain the result. If ${\Delta}K$ is due to the difference in $K_{\rm orb}$, the slope of $K^{\rm O(3)}$ vs. $K^{\rm O(2)}$ plot must not be dependent on the temperature. However, as seen in Supplementary Figure S3, the slope changes below $T_{\rm nem}$\cite{SM}.

Then, we consider the electronic nematicity in the pseudogap phase\cite{Kivelson1998}. The magnetic field is applied along the $c$-axis in our study, which does not introduce an additional symmetry breaking force.
Thus, we can assess that ${{\chi }^{\text{O(2)}}}>{{\chi }^{\text{O(3)}}}$. The inequivalence of magnetic susceptibility is most likely from the spin part because the averaging $K_{\rm orb}$ of O(2) and O(3) is only -0.008 $\%$\cite{Mangelschots1992}, which is much smaller than $K_{\rm spin}$, and should be $T$-independent. Thus, the local spin susceptibility is larger for O(2) than for O(3), 
meaning that the electronic $C_4$ symmetry is broken within each CuO$_2$ unit cell. 
This is consistent with the observation obtained via scanning tunneling microscopy (STM) in Bi$_2$Sr$_2$CaCu$_2$O$_{8+\delta}$ (Bi2212), where the electronic states along the $a$- and $b$-axis are found to be inequivalent \cite{Lawler2010}.
The $C_4$ symmetry breaking in the pseudogap phase has been observed through various bulk measurements in YBa$_2$Cu$_3$O$_y$\cite{Hinkov2008,Daou2010,Sato2017}. In particular, a similar observation indicating that ${{\chi }_{aa}}>{{\chi }_{bb}}$ has been reported by a recent torque measurement\cite{Sato2017}. Because O(2) lies in the Cu-O-Cu bond along the $a$-axis, our data show that the origin of the anisotropy of magnetic susceptibility is the inequivalence of the local magnetic susceptibility of the O(2) and O(3) sites.
The obtained $T_{\rm nem}$ is plotted in Fig. \ref{phase}, which enriches the global phase diagram of the YBCO systems\cite{Sato2017}. In passing, we note that the spin susceptibility is expected to be reduced below $T_c$ leading to a decrease of $\Delta K$, if the main contribution to it is from the spin part. However, due to the line broadening from the formation of vortices, $\Delta$$K$ cannot be obtained below 50 K\cite{SM,Machi99}. Therefore, studying the behavior of $\Delta$$K$ below $T_c$ is an important task in the future.

\subsection{Short-range CDW order}

\begin{figure}
\includegraphics[width= 8 cm]{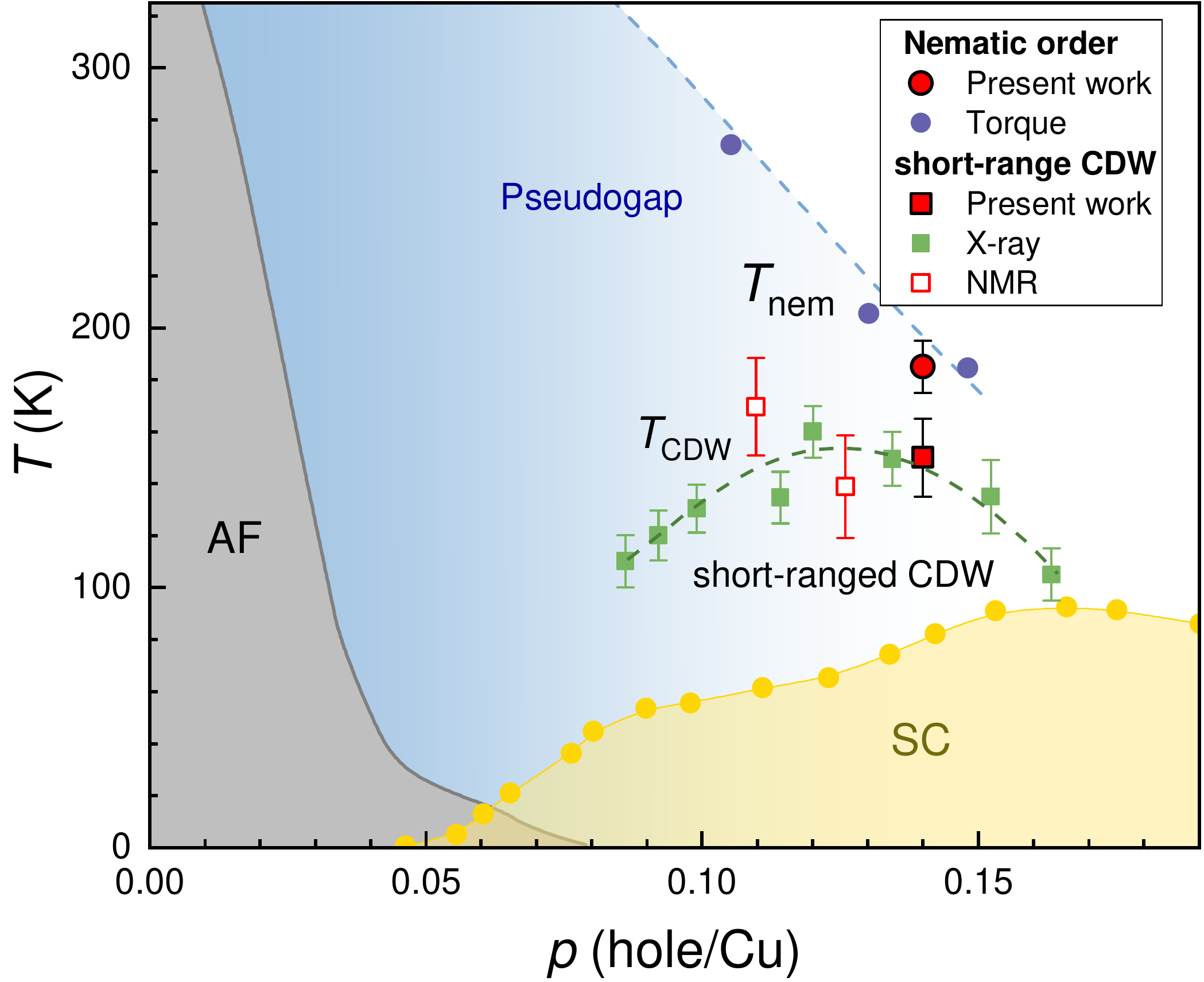}
\caption{ (Color online) The results obtained in this work are plotted in the global phase diagram. The red circle is the onset temperature $T_{\rm nem}$ of nematicity in YBa$_2$Cu$_4$O$_8$ determined based on the emergence of $\Delta K$. The red square is the onset temperature $T_{\rm CDW}$ of short-range CDW in YBa$_2$Cu$_4$O$_8$ determined from the $T$-dependence of the $^{17}$O-NMR linewidth. 
The blue circle is the nematic transition temperature $T_{\rm nem}$ reported using torque magnetometry in YBa$_2$Cu$_3$O$_y$\cite{Sato2017}. The open red and solid green squares are $T_{\rm CDW}$ reported by NMR\cite{Wu2015} and through the resonant X-ray scattering measurements in YBa$_2$Cu$_3$O$_y$\cite{Huecker2014,Blanco-Canosa2014}, respectively. AF and SC denote the antiferromagnetic ordered and superconducting phase, respectively. }
\label{phase}
\end{figure}

Next, we search for the CDW order in YBa$_2$Cu$_4$O$_8$. In the charge-ordered state, both local spin susceptibility and EFG will become site-dependent because of charge modulation. When the local charge density changes, the magnetic and quadrupole shifts will also change\cite{Kawasaki2015}.
Therefore, in a long-range CDW state, the NMR lines will split\cite{Wu2013,Kawasaki17}. The splitting $\Delta$$f$ will be different for the high- and low-frequency satellite peaks because the change in the quadrupole shift will make the high- and low-frequency satellites move in opposite directions. For HF2 and LF2, the splitting will be ${{\Delta }_{\text{m}}}\text{+}2{{\Delta }_{\text{Q}}}$ and ${{\Delta }_{\text{m}}}\text{-}2{{\Delta }_{\text{Q}}}$, respectively. Here, ${{\Delta }_{\text{m}}}$ and ${{\Delta }_{\text{Q}}}$ are the splitting due to the Knight and quadrupole shifts, respectively. For a static short-range CDW state, spectrum broadening will occur instead of line splitting\cite{Li2016}. For HF2 and LF2, the broadening will be $\left|{{\delta }_{\text{m}}}+2{{\delta }_{\text{Q}}}\right|$ and $\left|{{\delta }_{\text{m}}}-2{{\delta }_{\text{Q}}}\right|$, respectively. Here, ${{\delta }_{\text{m}}}$ and ${{\delta }_{\text{Q}}}$ are the broadening due to the CDW order contributed by the magnetic and quadrupole parts, respectively. If ${{\delta }_{\text{m}}}$ and ${{\delta }_{\text{Q}}}$ are comparable, a similar $T$-dependent broadening but with a distinct magnitude can be observed for HF2 and LF2 upon cooling.
The doping-induced inhomogeneity can result in ${{\delta }_{\text{m}}}$ and ${{\delta }_{\text{Q}}}$; however, they must be $T$-independent and appear at high temperatures.


\begin{figure}
\includegraphics[width= 8 cm]{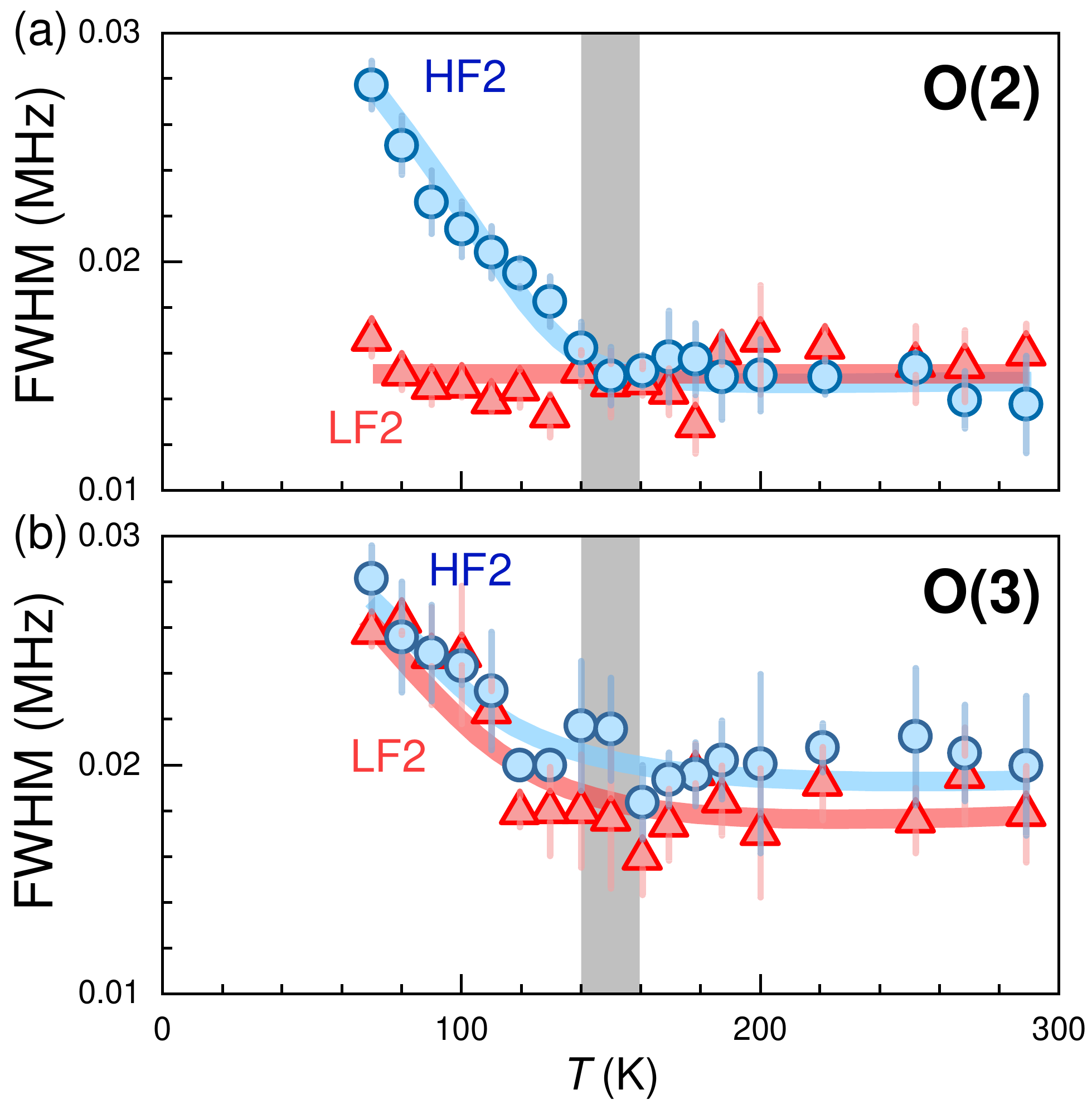}
\caption{(Color online) Evidence of the short-range charge density wave order. Panels (a) and (b) show the temperature dependence of the full width at half maximum (FWHM) of HF2 and LF2 for O(2) and O(3), which was measured in a magnetic field of 16.05 T. Solid lines are the guide to the eye.
}
\label{width}
\end{figure}

Figure \ref{width} panels (a) and (b) show the comparison of FWHM of HF2 and LF2 for the O(2) and O(3) sites. At high temperatures, the FWHM of HF2 and LF2 is identical for both the O(2) and O(3) sites, implying small doping inhomogeneity in our sample. However, below $T \sim$ 150 K, the FWHM of O(2) HF2 begins to increase, whereas the FWHM of LF2 is almost $T$-independent. The distinct $T$-dependent behavior of HF2 and LF2 implies that ${{\delta }_{\text{m}}}$ and ${{\delta }_{\text{Q}}}$ start to increase below $T \sim$ 150 K. This clearly demonstrates that a static short-range CDW order occurs because the time scale of NMR spectroscopy is in MHz. Moreover, as discussed above, if ${{\delta }_{\text{LF2}}}$ = $\left|{{\delta }_{\text{m}}}-2{{\delta }_{\text{Q}}}\right|$, then the $T$-independent behavior of the FWHM of LF2 indicates that ${{\delta }_{\text{m}}}$ is approximately twice as large as ${{\delta }_{\text{Q}}}$.
For O(3), LF2 and HF2 are broadened with a similar amplitude upon cooling below $T_{\rm CDW}$, which implies that either ${{\delta }_{\text{m}}}$ or ${{\delta }_{\text{Q}}}$ is very small. Because the broadening at HF1 is much smaller than that at HF2\cite{SM}, the $T$-dependent FWHM for O(3) suggests that ${{\delta }_{\text{m}}}$ is almost zero. Therefore, the ${{\delta }_{\text{m}}}$ of O(2) is much larger than that of O(3). A similar behavior has also been observed in YBa$_2$Cu$_3$O$_y$, which suggests that the observed CDW is uniaxial or biaxial with $d$-symmetry\cite{Wu2015}.
Overall, our results indicate that the same short-range nematic CDW order appears in YBa$_2$Cu$_4$O$_8$ below $T_{\rm CDW} \sim$ 150 K, as summarized in Fig. \ref{phase}, which coincides with the $T^{\ast}_{\rm max}$ obtained at which 1/$T_1$$T$ shows the maximum, as has been previously observed in YBa$_2$Cu$_3$O$_y$\cite{Huecker2014}. Our results further suggest that Fermi surface reconstruction by quantum oscillations can be attributed to the CDW order\cite{Yelland2008,Tan2015}. Noting that there are much less oxygen vacancies in the chains in our case, we speculate that the observed static short-range CDW is due to the small amount of defects in the CuO$_2$ planes that pin  the CDW fluctuations.



\section{Discussion}\label{section3}

As observed in case of YBa$_2$Cu$_3$O$_y$, the onset temperature of the nematic order of YBa$_2$Cu$_4$O$_8$ is very close to the pseudogap temperature $T^{\ast}$ but higher than $T_{\rm CDW}$, as shown in Fig. \ref{phase}, suggesting that such intra-unit-cell electronic nematicity can exist by itself. This also suggests that nematicity may be directly related to the pseudogap. Therefore, nematicity is essential to understand the nature of pseudogap. Our results provide new information for the theoretical modeling of nematicity inside the pseudogap phase.
Several proposals relating nematicity to various competing orders in the pseudogap phase have been introduced previously\cite{Kivelson1998,Nie2014,Tsuchiizu2018,Lee2018,Orth2019,Varma2009}. The first proposal is the melting of the uniaxial-density wave order such as stripe order\cite{Kivelson1998}. In our study, an intrinsic short-range CDW order, which exhibited different charge modulation at O(2) and O(3) sites, is observed at a lower temperature than $T_{\rm nem}$,
Another proposal is that nematicity is linked to antiferromagnetic fluctuations near the Mott phase\cite{Tsuchiizu2018,Orth2019}. In this scenario, the local charge densities at O(2) and O(3) are predicted to be inequivalent\cite{Tsuchiizu2018,Orth2019}. However, this contradicts our observation that the difference in EFG at O(2) and O(3) remains unchanged below $T^\ast$, as shown in Fig. \ref{asy}.
Apparently, more theoretical studies have to be done in these regards.

\section{Conclusions}\label{section4}

In summary, we performed $^{17}$O-NMR measurements at two planar oxygen sites in YBa$_2$Cu$_4$O$_8$. We observed that the difference in the Knight shift of the two oxygen sites begins to emerge below $T_{\rm nem} \simeq$ 185 K, which is very close to the pseudogap temperature $T^{\ast}$ and provides microscopic evidence for intra-unit-cell electronic nematicity. Further, the inequivalence of the quadrupole frequency $\nu_c$ between two oxygen sites is $T$-independent, indicating that electronic nematicity is not directly related to the charge difference between two planar oxygens. At lower temperatures, the static short-range nematic CDW order can be observed below $T_{\rm CDW} \simeq$ 150 K, which implies that the observed nematicity may be related to the nematic CDW order.
Our work adds new insights to the global phase diagram of cuprates and provides important constraints for the theoretical modeling of nematic order in cuprates. We hope that these results will stimulate further experimental work on cuprates and Fe-based superconductors.

\begin{acknowledgments}
This work was supported by the National Natural Science Foundation of China (No. 11974405, No. 11674377, No. 11634015), MOST grants (No. 2016YFA0300502 and No. 2017YFA0302904), and the Strategic Priority Research Program of the Chinese Academy of Sciences (Grant No. XDB33010100).
\end{acknowledgments}


\begin{references}


\bibitem{Fradkin2010} E. Fradkin, S. A. Kivelson, M. J. Lawler, J. P. Eisenstein, and A. P. Mackenzie, \href{https://doi.org/10.1146/annurev-conmatphys-070909-103925}{Annu. Rev. Condens. Matter Phys. {\bf 1}, 153-178 (2010).}

\bibitem{Keimer2010} B. Keimer, S. A. Kivelson, M. R. Norman, S. Uchida, and J. Zaanen,
\href{https://doi.org/10.1038/nature14165}{Nature \textbf{518}, 179-186 (2015).}

\bibitem{Lawler2010} M. J. Lawler, K. Fujita, J. Lee, A. R. Schmidt, Y. Kohsaka, C. K. Kim, H. Eisaki, S. Uchida, J. C. Davis, J. P. Sethna, and E.-A. Kim,
\href{https://doi.org/10.1038/nature09169}{Nature \textbf{466}, 347-351(2010).}

\bibitem{Mesaros2011} A. Mesaros, K. Fujita, H. Eisaki, S. Uchida, J. C. Davis, S. Sachdev, J. Zaanen, M. J. Lawler, and E.-A. Kim,
\href{https://doi.org/10.1126/science.1201082}{Science \textbf{333}, 426-430 (2011).}


\bibitem{Wu2017} J. Wu, A. T. Bollinger, X. He, and I. Bo\v{z}ovi\'{c},  \href{https://doi.org/10.1038/nature23290}{Nature \textbf{547}, 432-435 (2017).}

\bibitem{Auvray2019} N. Auvray, B. Loret, S. Benhabib, M. Cazayous, R. D. Zhong, J. Schneeloch, G. D. Gu, A. Forget, D. Colson, I. Paul, A. Sacuto, and Y. Gallais,  \href{https://doi.org/10.1038/s41467-019-12940-w}{Nat. Commun. \textbf{10}, 5209 (2019).}






\bibitem{Kohsaka2007} Y. Kohsaka, C. Taylor, K. Fujita, A. Schmidt, C. Lupien, T. Hanaguri, M. Azuma, M. Takano, H. Eisaki, H. Takagi, S. Uchida, and J. C. Davis,
\href{https://doi.org/10.1126/science.1138584}{Science \textbf{315}, 1380-1385 (2007).}



\bibitem{Hinkov2008}V. Hinkov, D. Haug, B. Fauqu\'{e}, P. Bourges, Y. Sidis, A. Ivanov, C. Bernhard, C. T. Lin, B. Keimer,
\href{https://doi.org/10.1126/science.1152309}{Science \textbf{319}, 597-600 (2008).}


\bibitem{Daou2010} R. Daou, J. Chang, D. LeBoeuf, O. Cyr-Choini\`{e}re, F. Lalibert\'{e}, N. Doiron-Leyraud, B. J. Ramshaw, R. Liang, D. A. Bonn, W. N. Hardy and L. Taillefer,
 \href{https://doi.org/10.1038/nature08716}{Nature \textbf{463}, 519-522 (2010).}



\bibitem{Sato2017} Y. Sato, S. Kasahara, H. Murayama, Y. Kasahara, E.-G. Moon, T. Nishizaki, T. Loew, J. Porras, B. Keimer, T. Shibauchi, and Y. Matsuda,  \href{https://doi.org/10.1038/nphys4205}{Nat. Phys. \textbf{13}, 1074-1078 (2017).}




\bibitem{Shekhter2013} A. Shekhter, B. J. Ramshaw, R. Liang, W. N. Hardy, D. A. Bonn, F. F. Balakirev, R. D. McDonald, J. B. Betts, S. C. Riggs, and A. Migliori,  \href{https://doi.org/10.1038/nature12165}{Nature \textbf{498}, 75-77 (2013).}

\bibitem{Badoux2016} S.Badoux, W. Tabis, F. Lalibert\'{e}, G. Grissonnanche, B. Vignolle, D. Vignolles, J. B\'{e}ard, D. A. Bonn, W. N. Hardy, R. Liang, N. Doiron-Leyraud, L. Taillefer, and C. Proust,  \href{https://doi.org/10.1038/nature16983}{Nature \textbf{531}, 210-214 (2016).}


\bibitem{Lederer2017}
 S. Lederer, Y. Schattner, E. Berg, and S. A. Kivelson, \href{https://doi.org/10.1073/pnas.1620651114}{Proc. Natl. Acad. Sci. U. S. A. {\bf 114}, 4905-4910 (2017).}



\bibitem{Kee09} H.-Y. Kee and D. Podolsky,  \href{https://doi.org/10.1209/0295-5075/86/57005}{EPL \textbf{86}, 57005 (2009).}

\bibitem{Wu2011} T. Wu, H. Mayaﬀre, S. Kr\"{a}mer, M. Horvati\'{c}, C. Berthier, W. N. Hardy, R. Liang, D. A. Bonn, and M.-H. Julien, \href{http://dx.doi.org/10.1038/nature10345}{Nature {\bf 477}, 191-194 (2011).}

\bibitem{Ghiringhelli2012} G. Ghiringhelli, M. Le Tacon, M. Minola, S. Blanco-Canosa, C. Mazzoli, N. B. Brookes, G. M. De Luca, A. Frano, D. G. Hawthorn, F. He, T. Loew, M. Moretti Sala, D. C. Peets, M. Salluzzo, E. Schierle, R. Sutarto, G. A. Sawatzky, E. Weschke, B. Keimer, and L. Braicovich,
 \href{https://doi.org/10.1126/science.1223532}{Science {\bf 337}, 821-825 (2012)}.



\bibitem{Gerber2015} S. Gerber, H. Jang, H. Nojiri, S. Matsuzawa, H. Yasumura, D. A. Bonn, R. Liang, W. N. Hardy, Z. Islam, A. Mehta, S. Song, M. Sikorski, D. Stefanescu, Y. Feng, S. A. Kivelson, T. P. Devereaux, Z.-X. Shen, C.-C. Kao, W.-S. Lee, D. Zhu, and J.-S. Lee, \href{https://doi.org/10.1126/science.aac6257}{Science {\bf 350}, 949-952 (2015).}

\bibitem{Chang2016}  J. Chang, E. Blackburn, O. Ivashko, A. T. Holmes, N. B. Christensen, M. H\"{u}cker, R. Liang, D. A. Bonn, W. N. Hardy, U. R\"{u}tt, M. v. Zimmermann, E. M. Forgan, and S. M. Hayden,  \href{https://doi.org/10.1038/ncomms11494}{Nat. Commun. {\bf 7}, 11494 (2016)}.



\bibitem{Fu2012} M. Fu, D. A. Torchetti, T. Imai, F. L. Ning, J.-Q. Yan, and A. S. Sefat,  \href{https://doi.org/10.1103/PhysRevLett.109.247001}{Phys. Rev. Lett. \textbf{109}, 247001 (2012).}

\bibitem{Baek2015} S-H. Baek, D. V. Efremov, J. M. Ok, J. S. Kim, J. van den Brink, and B. B\"{u}chner, \href{https://doi.org/10.1038/nmat4138}{Nat. Mater. \textbf{14}, 210-214 (2015).}

\bibitem{Zhou2016} R. Zhou, L. Y. Xing, X. C. Wang, C. Q. Jin, and G.-q. Zheng, \href{https://doi.org/10.1103/PhysRevB.93.060502}{Phys. Rev. B \textbf{93}, 060502(R) (2016).}

\bibitem{Iye2015} T. Iye, M.-H. Julien, H. Mayaﬀre, M. Horvati\'{c}, C. Berthier, K. Ishida, H. Ikeda, S. Kasahara, T. Shibauchi, and Y. Matsuda,  \href{https://doi.org/10.7566/JPSJ.84.043705}{J. Phys. Soc. Jpn. \textbf{84}, 043705 (2015).}

\bibitem{Dioguardi2016} A. P. Dioguardi, T. Kissikov, C. H. Lin, K. R. Shirer, M. M. Lawson, H.-J. Grafe, J.-H. Chu, I. R. Fisher, R. M. Fernandes, and N. J. Curro,  \href{https://doi.org/10.1103/PhysRevLett.116.107202}{Phys. Rev. Lett. \textbf{116}, 107202 (2016).}


\bibitem{Wu2016} T. Wu, R. Zhou, M. Hirata, I. Vinograd, H. Mayaffre, R. Liang, W. N. Hardy, D. A. Bonn, T. Loew, J. Porras, D. Haug, C. T. Lin, V. Hinkov, B. Keimer, and M.-H. Julien \href{https://link.aps.org/doi/10.1103/PhysRevB.93.134518} {Phys. Rev. B {\bf 93}, 134518 (2016).}



\bibitem{Kodama1990} Y. Kodama, Y. Yamada, N. Murayama, M. Awano, and T. Matsumoto,  \href{https://doi.org/10.1007/978-4-431-68141-0_88}{Advances in Superconductivity III, pp. 399-402 (1990).}

\bibitem{Zheng1992} G.-q. Zheng, Y. Kitaoka, K. Asayama, Y. Kodama, and Y. Yamada,  \href{https://doi.org/10.1016/0921-4534(92)90882-D}{Physica C \textbf{193}, 154-162 (1992).}


\bibitem{SM} See Supplemental Material for additional data and analysis.

\bibitem{Liang2006} R. Liang, D. A. Bonn, and W. N. Hardy,  \href{https://doi.org/10.1103/PhysRevB.73.180505}{Phys. Rev. B \textbf{73}, 180505 (2006).}




\bibitem{Vinograd2019}  I. Vinograd, Ph. D. thesis, Universit\'{e} Grenoble Alpes, France, 2018.  \href{https://tel.archives-ouvertes.fr/tel-02098091}{tel-02098091.}




\bibitem{Tomeno1994} I. Tomeno, T. Machi, K. Tai, N. Koshizuka, S. Kambe, A. Hayashi, Y. Ueda, and H. Yasuoka, \href{https://doi.org/10.1103/PhysRevB.49.15327}{Phys. Rev. B \textbf{49}, 15327 (1994).}

\bibitem{Zheng1999}G.-q. Zheng, W. G. Clark, Y. Kitaoka, K. Asayama, Y. Kodama, P. Kuhns, and W. G. Moulton,
\href{https://doi.org/10.1103/PhysRevB.60.R9947}{Phys. Rev. B \textbf{60}, R9947(R) (1999).}

\bibitem{Bankay1994} M. Bankay, M. Mali, J. Roos, and D. Brinkmann, \href{https://doi.org/10.1103/PhysRevB.50.6416}{Phys. Rev. B \textbf{50}, 6416 (1994).}

\bibitem{Hussey1997} N. E. Hussey, K. Nozawa, H. Takagi, S. Adachi, and K. Tanabe,
\href{https://doi.org/10.1103/PhysRevB.56.R11423}{Phys. Rev. B \textbf{56}, R11423(R) (1997).}

\bibitem{Alexandrov1990} O. V. Alexandrov, M. Fran\c{c}ois, T. Grafa, and K. Yvon,
\href{https://doi.org/10.1016/0921-4534(90)90228-7}{Physica C \textbf{170}, 56-58 (1990).}

\bibitem{Kivelson1998} S. A. Kivelson, E. Fradkin, and V. J. Emery,
\href{https://doi.org/10.1038/31177}{Nature \textbf{393}, 550-553 (1998).}

\bibitem{Mangelschots1992} I. Mangelschots, M. Mali, J. Roos, D. Brinkmann, S. Rusiecki, J. Karpinski, and E. Kaldis,
\href{https://doi.org/10.1016/S0921-4534(05)80005-X}{Physica C \textbf{194}, 277-286 (1992).}

\bibitem{Machi99}T. Machi, N. Koshizuka, and H. Yasuoka,
\href{https://doi.org/10.1007/978-4-431-66874-9_48}{Advances in Superconductivity \textbf{XI}, pp.227-230 (1999).}

\bibitem{Kawasaki2015} S. Kawasaki, Y. Tani, T. Mabuchi, K. Kudo, Y. Nishikubo, D. Mitsuoka, M. Nohara, and G.-q. Zheng, \href{https://doi.org/10.1103/PhysRevB.91.060510}{Phys. Rev. B \textbf{91}, 060510(R) (2015).}

\bibitem{Wu2013} T. Wu, H. Mayaﬀre, S. Kr\"{a}mer, M. Horvati\'{c}, C. Berthier, P. L. Kuhns, A. P. Reyes, R. Liang, W. N. Hardy, D. A. Bonn, and M. H. Julien, \href{https://doi.org/10.1038/ncomms3113} {Nat. Commun. {\bf 4}, 2113 (2013).}

\bibitem{Kawasaki17} S. Kawasaki, Z. Li, M. Kitahashi, C. T. Lin, P. L. Kuhns, A. P. Reyes, and G.-q. Zheng,  \href{https://doi.org/10.1038/s41467-017-01465-9}{Nat. Commun. \textbf{8}, 1267 (2017).}

\bibitem{Li2016} Z. Li, W. H. Jiao, G. H. Cao, and G.-q. Zheng, \href{https://doi.org/10.1103/PhysRevB.94.174511}{Phys. Rev. B 94, 174511 (2016).}

\bibitem{Wu2015}T. Wu, H. Mayaﬀre, S. Kr\"{a}mer, M. Horvati\'{c}, C. Berthier, W. N. Hardy, R. Liang, D. A. Bonn, and M.-H. Julien,     \href{https://doi.org/10.1038/ncomms7438} {Nat. Commun. {\bf 6}, 6438 (2015).}

\bibitem{Huecker2014} M. H\"{u}cker, N. B. Christensen, A. T. Holmes, E. Blackburn, E. M. Forgan, R. Liang, D. A. Bonn, W. N. Hardy, O. Gutowski, M. v. Zimmermann, S. M. Hayden, and J. Chang,  \href{http://link.aps.org/doi/10.1103/PhysRevB.90.054514} {Phys. Rev. B~{\bf 90}, 054514 (2014).}

\bibitem{Blanco-Canosa2014}
S. Blanco-Canosa, A. Frano, E. Schierle, J. Porras, T. Loew, M. Minola, M. Bluschke, E. Weschke, B. Keimer, and M. Le Tacon,  \href{http://link.aps.org/doi/10.1103/PhysRevB.90.054513} {Phys. Rev. B~{\bf 90}, 054513 (2014).}

\bibitem{Yelland2008} E. A. Yelland, J. Singleton, C. H. Mielke, N. Harrison, F. F. Balakirev, B. Dabrowski, and J. R. Cooper, \href{https://doi.org/10.1103/PhysRevLett.100.047003}{Phys. Rev. Lett. \textbf{100}, 047003 (2008).}

\bibitem{Tan2015}
B. S. Tan, N. Harrison, Z. Zhu, F. Balakirev, B. J. Ramshaw, A. Srivastava, S. A. Sabok-Sayr, B. Dabrowski, G. G. Lonzarich, and S. E. Sebastian,  \href{https://doi.org/10.1073/pnas.1504164112}{Proc. Natl. Acad. Sci. U. S. A. {\bf 112}, 9568-9572 (2015).}



\bibitem{Nie2014} L. Nie, G. Tarjus, and S. A. Kivelson,  \href{https://doi.org/10.1073/pnas.1406019111}{Proc. Natl. Acad. Sci. U. S. A. {\bf 111}, 7980-7985 (2014).}

\bibitem{Tsuchiizu2018} M. Tsuchiizu, K. Kawaguchi, Y. Yamakawa, and H. Kontani, \href{https://doi.org/10.1103/PhysRevB.97.165131}{Phys. Rev B \textbf{97}, 165131 (2018).}

\bibitem{Lee2018}S. Lee, J. Jung, A. Go, and E.-G. Moon,  \href{https://arxiv.org/abs/1803.00578v1}{arXiv:1803.00578v1 (2018).}

\bibitem{Orth2019}P. P. Orth, B. Jeevanesan, R. M. Fernandes, and J. Schmalian, \href{https://doi.org/10.1038/s41535-018-0143-y}{NPJ Quantum Materials \textbf{4}, 4 (2019).}

\bibitem{Varma2009}A. Shekhter and C. M. Varma, \href{https://doi.org/10.1103/PhysRevB.80.214501}{Phys. Rev. B \textbf{80}, 214501 (2009).}







\end{references}

\end{document}